# Thermodynamics without Time

Francesca Vidotto

*Department of Physics and Astronomy, Department of Philosophy, and Rotman Institute, Western University, N6A5B7 London, Ontario, Canada.*

*Instituto de Estructura de la Materia, IEM-CSIC, C/ Serrano 121, 28006 Madrid, Spain.*

**Abstract.** Our fundamental theories, i.e., the quantum theory and general relativity, are invariant under time reversal. Only when we treat system from the point of view of thermodynamics, i.e., averaging between many subsystem components, an arrow of time emerges. The relation between thermodynamic and the quantum theory has been fertile, deeply explored and still a source of new investigations. The relation between the quantum theory and gravity, while it has not yet brought an established theory of quantum gravity, has certainly sparkled in depth analysis and tentative new theories. On the other hand, the connection between gravity and thermodynamics is less investigated and more puzzling. I review a selection of results in covariant thermodynamics, such as the construction of a covariant notion of thermal equilibrium by considering tripartite systems. I discuss how such construction requires a relational take on thermodynamics, similarly of what happens in the quantum theory and in gravity.

Thermodynamics developed on the basis of the classical pre-relativistic notions of absolute space and absolute time. With the revolution brought about by general relativity, these absolute notions have been understood not to be general enough to describe the entire world accessible to our experience, and a novel way conceptualizing space and time has developed. This development, and especially the revision of the notion of time, has an impact on the foundation of the thermodynamics. It raises the question: to what extent can thermodynamics be extended or modified to remain applicable in regimes where the classical, pre-relativistic conception of absolute time lose its coherence?

This question is only partially addressed in the theoretical physics literature. The extension of thermodynamics to special relativity has been developed by numerous authors, and is not particularly troublesome, besides perhaps on terminological disputes regarding how temperature transforms under Lorentz transformation. The extension of thermodynamics to the case of the thermal behaviour of material systems on a fixed gravitational field (the gravitational analog of the thermodynamics of charged matter in a given electromagnetic field) is quite understood as well (Tolman 1934).

What instead remains largely open is the thermodynamics of the gravitational field itself (the gravitational analog of the thermodynamics of the electromagnetic field, namely the black-body radiation problem). In nature, indeed, there is no reason for thermal energy not to dissipate also into gravitational degrees of freedom. The very spacetime geometry in which we live is thus a dynamical field with thermal properties. For these, we do not have yet a theoretical understanding. Part of the difficulty lies in the fact that gravity is attractive and long-range. Objects that interact gravitationally, even in the non-relativistic approximation, are challenging to describe with standard techniques. But, the root of the difficulty is much deeper than this, as discussed below. In the physics literature, the thermodynamics of the gravitational field is largely treated haphazardly in special cases only, like for black holes, in the complete lack of a foundation or a general theory.

Here I illustrate the main reasons underpinning the current confused state in the foundation of relativistic thermodynamics, and some ideas that have been put forward to resolve them. The core of the problem is the fact that the notion of absolute time used in the foundation of classical thermodynamics is not anymore available in general relativity. Therefore, I first briefly recall the reasons why this notion is not available in the context of relativistic gravitation (Section 1). I illustrate then a specific physical phenomenon, the Tolman effect, that clearly shows that conventional thermodynamics needs a foundational revision and provides several useful hints for possible alternative foundational ideas (Section 2). I then discuss the notion of thermal time, which could be a key idea pointing to a possible solution of the problem (Section 3) and the peculiar way in which the notion of equilibrium could be recovered in a relativistic context, not a as a property or two systems, but rather a property of three systems (Section 4). I close in Section 5 with some general remarks on the conceptual novelties needed for gravitational thermodynamics.

## 1. Notions of Time

Time enters in both the first and the second principles of thermodynamics: energy is conserved in time and entropy does not decrease in time. When thermodynamics was developed, time was understood as Newton's absolute time. This conception is characterized by its uniformity in space —meaning it is the same at all locations—and in time. This is a key aspect that distinguishes it from the time of general relativity, which is dynamical, affected by matter, and which flows at different rates in different locations, as witnessed by the fact that in a gravitational field in gener-al two clocks that meet twice measure two different amounts of time lapsed between the two encounters.

It is perhaps interesting to observe, here, that the Newtonian notion of time, so ingrained in our modern way of thinking, is not necessarily the most primitively intuitive, nor was common before Newton. Aristotle defined time as a measure of change and such way of understanding time remained dominant until Newton. If time is defined as a measure of change, it does not (unlikely Newtonian time) lapse when nothing changes. This definition of time, as we shall see, remains consistent within relativistic physics, it is compatible with a relational notion of change through the use of clocks, and even gets a very concrete physical interpretation in the properties of thermal time, illustrated below.

In general relativity, the notion of time is radically altered in a number of ways. First, time cannot be separated from space; it is conceived as an aspect of the unified entity called *spacetime*. The radical conceptual novelty introduced by Einstein with general relativity is the identification of this spacetime with the gravitational field. Since all fields in physics have a dynamic nature, this time is also in a sense dynamical. The rate at which it lapses (as defined by a clock) is affected by this dynamics. For instance, it is modified by the vicinity of masses.

In this rich conceptual scheme, the Newtonian monolithic notion of time break us in a variety of notions, only loosely connected to one another: coordinate time, proper time along world line, clock time, asymptotic time… In the non-relativistic flat-space limit, all these notions agree, but in general they are quite distinct notions, none of which retains the full properties of Newtonian time. Newtonian time is today understood as an approximation: in reality, there are no preferential variables that constitute an absolute time; at most there are particular physical situations where one variable plays the role of Newtonian time, but with a limited scope.

The Hamiltonian formulation of the theory makes this situation particularly vivid. A non-relativistic system admit a natural Hamiltonian formulation in which, in phase space, the evolution of the system in time is governed by (is the flux of) a Hamiltonian. Not so for general relativistic systems. These can only be naturally formulated in a generalized Hamiltonian formalism, where the dynamics is expressed by a constraint on the phase space: on the subspace where it vanishes, the constraint generates (its flux defines) a relation between variables.

If dynamics is understood as a relation between variables, then all variables are a priori on the same footing and many different clocks can play the role of independent variable that Newtonian time used to play. This relational understanding of temporality is perfectly fine for describing the world, and in fact general relativity does describe the world very well. But how do we do thermodynamics in a theory that refuses to provide us a preferred time variable?

Consider in particular the statistical underpinning of thermodynamics. We understand thermodynamic states as statistical distributions of micro-states. In a gravitational context, this appears to imply that a thermodynamic state of the gravitational field (say the gravitational analog of the electromagnetic black-body distribution) is actually a statistical mixture of spacetimes. In each of these clocks, it moves at a different rate. So, how to identify the global time variable needed for classical thermodynamics? Thermal fluctuations in the gravitational field affect the very rate of the clocks. Extending statistical mechanics to effectively describe the thermal fluctuations of spacetime is an open problem in physics (Rovelli 2013, Chirco and Josset 2017). Even neglecting quantum phenomena.

Quantum physics introduces further complications. In the quantum theory, clocks become quantum systems themselves (Rovelli 2015). They can be in a superposition. Superposition of clocks implies superposition of the causal structures they define. Below I shall ignore any quantum phenomenon, and just focus on the foundation of thermodynamics and statistical mechanics in the classical (non-quantum) relativistic gravitational context.

Perhaps surprisingly, this is a problem whose solution is even further off than quantizing gravity! We have consistent and reasonable tentative theories of quantum gravity (Rovelli and Vidotto 2014), that have not been confirmed empirically, but have a clear conceptual foundation (Rovelli and Vidotto 2022) and might be right. We do not have a consistent and reasonable tentative theory of statistical gravity with a clear conceptual foundation.

## 2. Tolman-Ehrenfest effect

At the very core of thermodynamics, there is the notion of equilibrium. This notion is based on the observation that if two bodies, $A$ and $B$, at different temperature can exchange energy, they do so in such a way to converge to a common temperature. They equilibrate. At equilibrium, the temperature of the two bodies is the same: $T_A = T_B$. Some authors have emphasized this basic fact by giving the name of "Zeroth Principle of Thermodynamics". It is a basic fact stated at the beginning of any thermodynamics textbook, and it is very clear.

Except that in a gravitational field this is not true. Consider a concrete situation: take a vertical cylinder filled with gas or a fluid. Call the lower part of the cylinder $A$ and the upper part $B$, then move a thermometer at the bottom and at the top, and check whether it registers the same temperature. It does not. Rather, it measures a tiny temperature difference $\Delta T/T = gh/c^2$, where $g$ is the Galileo's gravitational acceleration ($\sim 9.8 m/s^2$), $h$ is the height difference and $c$ is the speed of light.

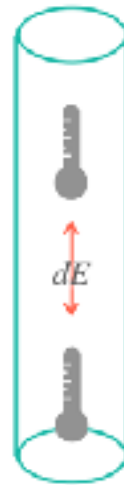

This phenomenon is called Tolman effect. It was first understood by Richard Tolman in 1930 in his paper "On the Weight of Heat and Thermal Equilibrium in General relativity" (Tolman 1930) and then generalized to wider contexts in collaboration with Ehrenfest (Tolman and Ehrenfest 1930). The phenomenon has never been directly measured because on Earth it is very small, but it has been derived in a variety of way and is universally considered credible.

It is a genuinely relativistic phenomenon (it disappears in the $c$ to infinity limit) and it is a genuine gravitational phenomenon: it disappears if the limit in which Newton constant (hence $g$) is taken to zero.

The derivations of the phenomenon are numerous, but none relies on first principles. Each of them is, so-to-say, a patchwork of different chunks of physics. Here is one: Temperature is inversely proportional to the derivative to the entropy with respect to the energy: $\frac{dS(E)}{dE} = \frac{1}{kT}$, where $\kappa$ is the Boltzmann constant. This allow us to understand temperature in terms of the micro-states of a given system, since the entropy $S(E)$ is the logarithm of the number of microstates $N(E)$ that have energy $E$. Say two bodies can exchange energy so that a small amount of energy $dE$ can exit one and the same amount of energy enter the other. When this happens the entropy of each changes. Say equilibrium is when the total entropy, which is the sum of the two, is at a maximum; this immediately implies that the two temperatures must be the same, because the two derivatives with respect to $dE$ must be the same.

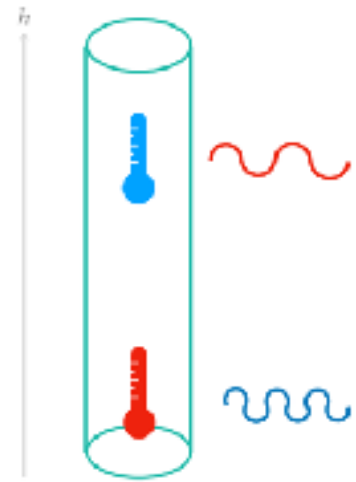

However, in a gravitational field the energy that exits one body is not the same as the energy that enters the other! The reason is that energy is like mass: it is attracted by gravity, hence it falls. In a relativistic context, that is, energy weights. Hence it acquires energy by falling, like any mass does. Hence if some energy $dE$ exits the upper body, the energy that enters the lower body would be larger. Hence the maximal entropy is not attained when the two derivatives are equal. The energy increase in falling by a bit of energy $dE$ can be easily estimated for instance thinking that it is carried by a photon. A photon is blue shifted in falling (this is a general relativistic effect) with a change in frequency $\Delta \nu / \nu = gh/c^2$. Since the energy of a photon is proportional to its frequency, so changes its energy in falling, and therefore the two temperatures have to be off by this same factor, in order for entropy to be maximal and entropy to be reached.

All this is clear and fully convincing. Yet, the conclusion is radical: in a gravitational field, equilibrium does not mean that temperature (as measured by the same thermometer) is uniform. The Zeroth principle of thermodynamics collapses.

A more precise statement of Tolman low is that in a static gravitational field described by a Newton potential $\Phi(x)$, equilibrium is given by

$$T_A\left(1 + \frac{\Phi(x_A)}{c^2}\right) = T_B\left(1 + \frac{\Phi(x_B)}{c^2}\right). \quad (1)$$

Equilibrium that is given by equality of the temperature at those positions multiplied by the redshift factor determined by the gravitational potential.

This observation can be extended (Chirco et al. 2013). In a generic gravitational field, we can talk about equilibrium only if there is a sense in which spacetime is stationary. This is when there is a time-like Killing field $\xi$. (A Killing field is a vector field along whose integral lines the metric remains the same.) In this case, the Tolman equilibrium equation above generalizes to $T|\xi| = const$, namely: what is constant at equilibrium is the product of the temperature times the norm of the Killing field.

At the light of this, one could of course define a notion of *proper temperature*, which is the temperature measured by a thermometer scaled by the red shift factor. This is the actual *temperature* whose uniformity signals equilibrium. But it is not the quantity measured by the devices we call thermometers.

The Tolman relation, thus, shows an important property of the temperature (measured by thermometers): temperature is defined in a way that depends on the local time flow. Since in relativity Newtonian time is replaced by multifinger time, that is, by clocks, temperature becomes dependent on these, hence depends on space. The exact way in which this works is illustrated in the next section.

In any case, the Tolman effect makes clear that the standard non-relativistic way of thinking about thermodynamics does not generalizes to relativistic gravity. However, it also points us towards an intriguing direction for a different foundation of thermodynamics. This is illustrated below.

## 3. Thermal Time

The concept of thermal time was introduced by Carlo Rovelli in 1993 (Rovelli 1993) and extended to the full context of quantum field by him in collaboration with Alain Connes (Connes and Rovelli 1994) using the Tomita-Takesaky modular theory. In this discussion, we are not interested in the extension to the quantum domain, hence we only briefly mention the definition, in order then to focus on the foundations of classical thermodynamics. Let's illustrate the notion of thermal time, first in the classical case.

Given any statistical state $\rho$, represented as a distribution over phase space, we can define its *thermal Hamiltonian* to be the $H_\rho = -ln\rho$ and thermal time $\tau$ as the parameter of the flow that the thermal Hamiltonian generates on phase space. For later convenience, let us rather define it as a dimensionless parameter by dividing it by $\hbar$. Here $\hbar$ is only used as a unit of action, so far with no physical reason. That is, given any observable function $A$, we consider the flow

$$\frac{dA(\tau)}{d\tau} = \hbar\{A, -ln\rho\}. \tag{2}$$

(In the general quantum case, the thermal time is the parameter of the Tomita-flow of the state, seen as a function on the algebra.) In the special case in which the state is the Boltzmann-Gibbs state $\rho \sim e^{-\beta H}$, where $\beta = 1/kT$, we have clearly that

$$\tau = \frac{k}{\hbar}Tt. \tag{3}$$

That is, in this case the thermal time is the usual time scaled by the temperature.

Now, let's see what the thermal time in the case of the Tolman effect is. Locally, it is proportional to clock time, but the proportionality constant changes with the temperature, namely, it changes from point to point. But so does, and precisely in the same manner, as shown by the Tolman low, the parameter that labels the flow generated by the Killing field. In other words, the thermal time is precisely the time flow along which the system is equilibrium.

This is already a nice result, but there is much more: temperature is the ratio between the thermal time (namely the time along which there is equilibrium) and the clock time! This surprising result suggests a very compelling interpretation of temperature: it is the ratio between thermal time and proper time. Equivalently, it is the speed at which clocks move, in the time along which there is equilibrium. Temperature is, we can say, the speed of time (Rovelli and Smerlak 2011). This will become clearer in a moment.

In fact, there is even more, because the thermal time can be given a very concrete physical interpretation. For this, fix a unit of action. For instance, in view of quantum theory and the Heisenberg principle, consider units of action of size $\hbar$. Accordingly, consider cells in phase space of size $\hbar$ per degree of freedom. Now consider a statistical state concentrated in one such phase space cell. Call $t_o$ the average time it takes this state, under the dynamical flow, to move from one such cells to the next one. The same quantity can be defined in quantum theory by asking what is the time $t_o$ it takes a quantum state $\psi(0)$ to evolve to a state $\psi(t_o)$ such that $|\langle \psi(0)|\psi(t_o)\rangle| \ll 1$. In Haggard and Rovelli (2013) this quantity is computed in the case of a thermal context, and shown to be universal and only depending on the temperature:

(4) $$t_o = \frac{\hbar}{kT}$$

This in turn implies that the thermal time lapsed along a motion is nothing else than the number of such elementary steps $t_o$ along such motion:

(5) $$\tau = \frac{t}{t_o} = \frac{k}{\hbar}Tt.$$

Let's summarize. The system jumps from cell to cell in phase space. If we take the cells being of size $\hbar$, the system on average jumps $\tau$ cells during the clock time t. Thermal time counts the jumps. Temperature is the (in natural units) the number of cells the system jumps per unit of clock time.

Notice how all fits with the Tolman phenomenology: thermal time is defined globally. Not so clock time, because its rate is affected locally by gravity. Hence the ratio between the two, namely temperature, changes locally.

And finally, most spectacularly: two systems are in equilibrium along a motion, in general, not if the temperature is the same, but if the number of elementary cells they have jumped along the motion is the same!

Expressing this in terms of quantum information, we discover that thermal time counts a number of states. Here, the fundamental quantum discreteness plays a central role: thermal time counts how many cells of size $\hbar$ are present in phase space.

This way of viewing equilibrium shifts the language from that of energy to that of information, in the sense of number of states. We can say that equilibrium is when the information flow between the two systems vanished, where the information flow is the difference between the number of states of one system "seen" by the other during the common motion. We can then switch to a fully informational language. The general covariant condition for equilibrium is given by a zero-information flux. The two subsystems are in interaction with one another: each system has access to the amount of information, in the sense of Shannon given by the number of states, of the other system it is exposed to, during a process. Equilibrium is achieved when the two systems "see" the same number of states of the other system, i.e., when the value of the thermal time coincides for the two systems. Hence, equilibration is a process that selects a preferred common time. Notice that, conceived in this manner, equilibrium is not defined among states: it is defined among *processes*.

This is a hint for generalizing thermodynamics. While energy is a concept that in general relativity becomes ambiguous, information, counting of jumps, might provide a viable way to express thermodynamical quantities in a general relativist context with no preferred time variable.

## 4. Equilibrium in (dynamical) general covariant systems

The previous two sections dealt with equilibrium of matter in a given gravitational field. They illustrated some concepts and ideas that play a role in relativistic thermodynamics in the presence of gravity, but they did not address the problem of the thermodynamics of the gravitational field.

Let's get back to the question of why this is difficult. For this, let's backtrack again to the statistical underpinning of thermodynamics. In standard thermodynamics, dynamics is given by a Hamiltonian on a phase space generating time evolution. Statistical mechanics can be derived from an assumption about a probability distribution on phase space invariant, under this dynamics.

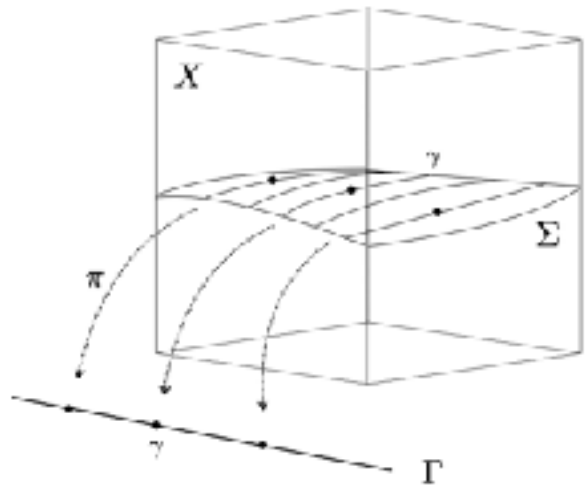

As mentioned, general relativity does not fit into this Hamiltonian formalism. Rather it fits into a *covariant* extension of this formalism, which I sketch below.

The (so called *extended*) phase space is again a symplectic space, but dynamics is not given by a Hamiltonian. Rather, it is given by a function $C$ on phase space called constraint. On the surface *C=0*, the motions generated by *C* itself are one-dimensional lines. Each of these lines is a motion. A motion can be arbitrarily parametrized with a parameter $t$, but this parameter is (indeed) arbitrary and in general does not correspond to anything physical or observable. Along the motion, functions $(A, B, C, ...)$ on phase space take values $(A(t), B(t), C(t), ...)$. Each such function $(A, B, C, ...)$ is associated with an operational procedure and can be observed. They are denoted "partial observables" (Rovelli 2002). What the theory predicts, and can be compared with experiments, is the relative dependence of these partial observable from one another, for instance $A(B)$, obtained solving $t$ away from $(A(t), B(t))$. The space of the motions is sometime called the *physical phase space*.

The connection between this general formalism and the special case of non-relativistic Hamiltonian mechanics is simple. The special case is obtained when one of the partial observables is identified with the Newtonian time *T* and *C* has the special form $C = p_T + H$, where $p_T$ is the variable canonical conjugate to *T*, and *H* is the usual Hamiltonian. In the general case, different partial variables can be identified with different kinds of clocks, as is the case in general relativity, where different real clocks measure different time lapses between the same couple of events.

The open question of general relativistic thermodynamics is how to define a useful statistical mechanics and thermodynamics, given only the generalized "covariant" Hamiltonian formalism sketched above. Some conceptual steps in this direction have been taken in (Chirco et al. 2016). I illustrate here the central idea. Consider a given partition of the system in two subsystems, that the extended phase space can be seen as the Cartesian product of two phase spaces. One of the two is interpreted as a clock the other is a physical system evolving in terms of the clock time. The resulting construction, illustrated below, depends on the choice of this partition.

In non-relativistic statistical mechanics, we can consider ergodic systems. For these, we can write an equilibrium measure $\nu$ on phase space by requiring that the the time average of an observable *f* along any motion $s(t)$ is equal to the phase space average of the observable computed with this measure

(6) $$\lim_{T \to +\infty} \frac{1}{2T} \int_{-T}^{+T} f(s(t))dt = \int_V f d\nu$$

This notion of ergodicity can be generalized to the relativistic case, as follows. First, choose a one form $d\theta$ on the phase space of the clock system. Then, given a finite motion $\gamma$, we can define the time average of an observable *f* along this motion by

(7)
$$\bar{f}(\theta,\gamma) \underset{def}{=} \frac{\int_\gamma f\theta}{\int_\gamma \theta}.$$

and say that a measure on the phase space is an equilibrium measure if this approaches the phase space of the observable with such measure, for every sufficiently long $\gamma$.

Now consider the case in which the system can be split into two subsystems, $S_a$ and $S_b$, such that the overall phase space is the Cartesian product of the two respective phase spaces and the constraint has the form $C = C_a + C_b$, where the first (rest second) term depends only on variables of the first (response second) system. Then, it is easy to see that this split determines a foliation of the constraint surface where the leaves are labelled by the value $I$ of $C_a = -C_b$, and each leaf, in turn, is the Cartesian product of one surface per system.

Interestingly, the total number of degrees of freedom of the system is the sum of the numbers of degrees of freedom of each subsystem —seen as an isolated constrained system defined by $C_a$ (rest $C_b$)—, plus one. The single additional degree of freedom is a relative variable connecting the two systems, and we can view $I$ as the value of the generator of its evolution. Now, say that we are interested in measuring partial observables of the subsystem $S_a$ only, as it evolves with respect to a clock defined by a variable in the subsystem, $S_b$ that is, by a one form $d\theta$ depending only on the variables of the second (clock) system. Then assuming ergodicity, it is proven in (Chirco et al 2016) that the equilibrium states on the phase space of the first system are independent from the choice of $d\theta$ and can be labelled by $I$ and written explicitly as

(8)
$$d\mu_{\Sigma_{I^a}^a} = \frac{\delta(C^a - I^a)d\mu_a}{\int \delta(C^a - I^a)d\mu_a},$$

where $\mu_a$ is the Liouville measure on the phase space of the first system. This result permits to define the full structure of statistical mechanics in a generally covariant system, on the sole basis of the choice of a (weakly- or non-interacting) clock subsystem.

Given now a split of the total system into *three* components, the above construction allows us to say when two of these are in equilibrium *with respect to the third*, namely, with respect to the clock system. This shows that in a generally covariant system, the notion of equilibrium is defined when three systems are given, not just two. Hence the subtitle of the paper Chirco et al (2016): *It Takes Three to Tango.* I refer the reader to that paper for rigorous derivations and extended details on the resulting form of thermodynamics.

# 5. Conclusions

A fundamental lesson of general relativity is the dissolution of time as a unique and absolute entity. It is nonetheless possible to generalize thermodynamics to the wider covariant context of relativistic gravity, where there is no preferred time. Above I have illustrated some ideas on the direction to go in order to do so. Here are some of the resulting possible lessons:

- In a covariant context, equilibrium between two systems needs to be defined *with respect to a third system: "It takes three to tango!"*

- Generically, every state defines its own thermal time. The thermal time of equilibrium states is a global notion. On a fixed background spacetime, it is globally proportional to the flow of the (time-like) Killing field along which equilibrium is established.

- Temperature is then not uniform, in general. Rather, it is defined locally, and it is the ratio between proper time and clock time. It is the *speed of time,* namely, the speed at which systems changed, measured in proper time.

- This rate of change can be defined as the rate at which the microphysics moves the system from a cell in space to the next one. Or, in quantum theory, a quantum states moves to an orthogonal state.

- At equilibrium, two systems may not have the same temperature. Rather, their rate of change is the same. This gets an informational interpretation by noticing that the number of states of one system that the other system is exposed to can be called quantity of information, or flux that can be acquired, or information flux. At equilibrium the net information flux (the difference between the fluxes in the two directions) vanishes.

- Equilibrium is then defined between processes, not states.

I close with some general observations. The use of the language of information theory should not be taken here as a reduction of everything to information. That would be the same mistake as going from the observation that energy is a very useful concept in numerous situations to the misleading and empty rhetoric of "everything is energy".

Still, this use of information language reinforces the idea that different conceptual domains often converge in forming a coherent vision of reality. When comparing theories with different metaphysical frameworks and the same empirical content, it is useful to consider the explanatory power they hold in light of their ability to offer a conceptual means for transitioning toward more fundamental and more explanatory theories. A second major observation is that *relational* thinking plays a major role in this context. Rovelli's work on the foundations of covariant thermodynamics is independent from his work on the interpretation of quantum mechanics and his introduction of the relational interpretation of quantum theory in the 90s (Rovelli 1996). Yet, in both cases, the interpretation of quantum mechanics and the formulation of relativistic gravitational statistical mechanics, problems are solved by bringing into view the underlying relational character of entities previously mistakenly taken as absolute (Vidotto 2022). Information plays a major role in both, as well.

The results presented here for covariant thermodynamics, much of which stems by Rovelli himself, show the influence of relational thinking that emerges from general relativity and (relational) quantum mechanics.

In the definition of equilibrium discussed above, a fundamental role is played by the partition of the system and the resulting relational degree of freedom that this partition singles out. It is this variable that ends up playing the role most similar to the Newtonian global time. Here a partition defines a relation (Rovelli 2014). Once again, the useful focus is on the interactions between systems, and on processes, rather than on the systems and their states.

Finally let me emphasize again that these results can be seen at best as the opening of a very intriguing research direction, certainly not yet as a complete theory. Also, the extension of these ideas to quantum theory is still entirely lacking.

Yet, it seems to me that this is where some of the more philosophically and conceptually interesting aspects of modern physics are nested: we have discovered that time works radically differently than in non-relativistic physics, and it is still taking long to integrate this profound discovery about the structure of really into our worldview. Rethinking physics without time in a fully relational way, prioritizing processes, may even require a deep conceptual discussion and, perhaps even openness to models of thought different from those traditionally employed in the development of Western science (Kohl 2012, Latten 2022).

## Acknowledgments

The author warmly thanks Carlo Rovelli, Matteo Smerlak, Hal Haggard, and Goffredo Chirco, for many discussions on this topic through the years. She is particularly grateful to Yichen Luo, Emily Adlam, Wayne Myrvold, Pascal Rodriguez, and all the researches who contributed to a Reading Group on Thermodynamics at Western University. FV's research at Western University is supported by the Canada Research Chairs Program, by the Natural Science and Engineering Council of Canada (NSERC) through the Discovery Grant "Loop Quantum Gravity: from Computation to Phenomenology", and by the John Templeton Foundation through the project ID# 62312 grant "The Quantum Information Structure of Spacetime" (QISS). FV acknowledges support from the Perimeter Institute for Theoretical Physics through its affiliation program. Research at Perimeter Institute is supported by the Government of Canada through Industry Canada and by the Province of Ontario through the Ministry of Economic Development and Innovation. FV acknowledges the Anishinaabek, Haudenosaunee, Lūnaapéewak, Attawandaron, and neutral people, on whose traditional lands Western University and the Perimeter Institute are located.